# *Bona fide* stochastic resonance under nonGaussian active fluctuations

Govind Paneru,[ab] Tsvi Tlusty,*[ab] and Hyuk Kyu Pak *[ab]

[a]Center for Soft and Living Matter, Institute for Basic Science (IBS), Ulsan 44919, Republic of Korea
[b]Department of Physics, Ulsan National Institute of Science and Technology, Ulsan 44919, Republic of Korea

We report on the experimental observation of stochastic resonance (SR) in a nonGaussian active bath without any periodic modulation. A Brownian particle hopping in a nanoscale double-well potential under the influence of nonGaussian correlated noise, with mean interval $\tau_P$ and correlation time $\tau_c$, shows a series of equally-spaced peaks in the residence time distribution at integral multiples of $\tau_P$. The strength of the first peak is found to be maximum when the mean residence time $\bar{\tau}_d$ matches the double condition, $4\tau_c \approx \tau_P \approx \bar{\tau}_d/2$, demonstrating a new type of *bona fide* SR. The experimental findings agree with a simple model that explains the emergence of SR without periodic modulation of the double-well potential. Additionally, we show that generic SR under periodic modulation, known to degrade in strongly correlated continuous noise, is recovered by the discrete nonGaussian kicks.

## Introduction

Stochastic resonance (SR) occurs when a weak *periodic* signal is enhanced in the presence of noise, and the enhancement shows resonant behavior as the noise is tuned [1]. A prototypical setting of SR is a Brownian particle hopping in a symmetric double-well potential under the influence of thermal noise. When periodic forcing is added, the double-well potential is asymmetrically tilted up and down, thereby sequentially raising and lowering the potential barriers. It may synchronize with the thermally-induced hopping process, as manifested by peaks in the residence time distribution (RTD) corresponding to the external force period. Maximum synchronization occurs when the thermally induced hopping rate is half the period of modulation. In this case, the first peak strength in the RTD assumes a maximum leading to *bona fide* SR[9]. SR is a universal phenomenon that has been explored in diverse fields ranging from climatology to biology [1-8]. The validity of SR as a *bona fide* resonance that attains maximal synchronization between periodic forcing and noise-induced hopping has been extensively discussed [9-13].

Can random signals also induce *bona fide* resonance? – Gammaitoni et al. [1, 14] demonstrated SR under a periodic signal with random amplitudes. However, whether fully random fluctuations, with random amplitude and random period, lead to *bona fide* resonance remains an open question. Here, we experimentally demonstrate SR of a Brownian particle in a double-well potential under nonGaussian active fluctuations having finite-amplitude active bursts of random amplitude that are arriving with an average period following a Poisson distribution. A direct implication of our finding is that SR is much more widespread than previously realized since it does not necessitate sequential forcing.

Such active baths with nonGaussian statistics have become a timely topic, as mounting evidence suggests they are prevalent in living systems. In the active baths around swimming bacteria [15-22] or in the cellular milieu [23-26], diffusion is governed by the coaction of uncorrelated thermal fluctuations of the solvent and correlated fluctuations induced by active components. While a common model of active baths has been the Active Ornstein–Uhlenbeck (AOU) noise [15, 27], theoretical studies showed that this Gaussian process suppresses SR as the active noise correlation time increases [28]. This result is considered counter-intuitive, as active noise generally enhances transport and diffusion, calling for a more realistic active bath model. Indeed, recent experiments provided evidence for nonGaussian noise processes in biological and artificial systems of active swimmers [17, 22, 29-31], as well as in living cells [32, 33] and glassy systems[34]. On the other hand, a recent theoretical study proposed that the non-Gaussian diffusion observed in a bacterial suspension can be explained in terms of a coloured Poisson process [35].

Here, we investigate the dynamics of a Brownian particle in a double-well potential under the influence of the exponentially correlated nonGaussian active noise. The nonGaussian active noise generates random-amplitude active bursts, decaying exponentially with correlation time $\tau_c$ and separated by discrete intervals with an average time $\tau_P$ following Poisson distribution (Fig. 1a). By adjusting the noise parameters, the active noise can switch between Gaussian and nonGaussian correlated noise, mimicking active baths with these statistics. We find that the particle position distribution exhibits two symmetric peaks that generally follow a Gaussian distribution. However, when $\tau_c < \tau_P$ and the active noise is stronger than the thermal noise, the peaks develop exponential side-tails, a signature of a nonGaussian active bath [17, 30, 36]. In this nonGaussian regime of the active noise, the RTD of the particle exhibits a series of exponentially decaying peaks at integral multiples of $\tau_P$. The strength of the first peak can be maximized by changing either $\tau_c$ or $\tau_P$, according to the resonance condition $4\tau_c \approx \tau_P \approx \bar{\tau}_d/2$, where $\bar{\tau}_d$ is the average residence time. This observation of the first peak maximum around the resonance value establishes the existence of

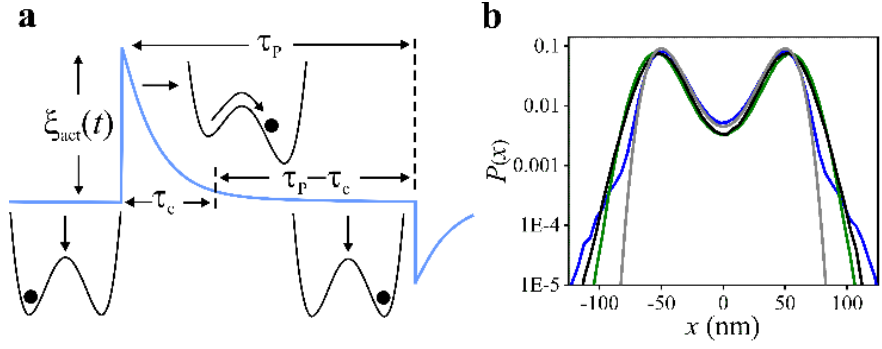

Fig. 1 (a) Illustration of SR in the presence of nonGaussian active noise without periodic modulation. The positive active burst $\xi_{act}(t)$ (blue curve) supplies energy to the particle in the double-well potential during a time $\tau_c$, effectively lowering the barrier height by lifting the left well. Subsequently, the strength of the active burst is significantly reduced during a time $\tau_P - \tau_c$, the left potential well is lowered back to its original position, and the thermal condition is recovered. (b) PDF of the particle position in the double-well potential, with $E_b/k_BT = 3$ and $x_m = 50$ nm, in the presence of active noise of fixed strength $f_{act} \approx 0.5$ pN and noise arrival interval $\tau_P \approx 28$ ms, for correlation times $\tau_c \approx 21$ ms (olive), 7 ms (black), 0.28 ms (blue, numerical result obtained by solving Eq. (1)). The gray curve is the theoretical PDF of the particle in a thermal bath alone, $P(x) \sim \exp(-V_{DW}(x)/k_BT)$.

*bona fide* SR *in active bath without any periodic modulation*, as the observed dynamics is driven solely by the nonGaussian active noise.

We further examine how the nonGaussian active noise affects the generic form of SR, *i.e.*, under a periodic force, which tilts the double-well potential up and down sequentially. We observe the suppression of the generic SR in the Gaussian regime of the active noise, a first experimental demonstration of the predicted degradation of SR in the presence of coloured noise [28]. Strikingly, the SR recovers in the nonGaussian regime, $\tau_P > \tau_c$. Overall, our results propose the correlated-nonGaussian noise as a strong generator of SR, with direct implications on stochastic processes in living systems.

## Results

**Active Bath Model.** We consider the one-dimensional motion of a Brownian particle in a symmetric double-well potential, $V_{DW}(x) = E_b[-2(x/x_m)^2 + (x/x_m)^4]$, where $x$ is the particle position, $\pm x_m$ are the potential minima, and $E_b$ is the barrier height in an active bath of temperature $T$.

The motion of the particle is described by the overdamped Langevin equation:

$$\gamma \frac{dx}{dt} = -\frac{\partial V_{DW}(x)}{\partial x} + \xi_{th} + \xi_{act}. \quad (1)$$

Here, the thermal noise $\xi_{th}$ is Gaussian white noise with zero mean and no memory, $\langle \xi_{th}(t)\xi_{th}(t') \rangle = 2\gamma^2 D \delta(t-t')$, where $\gamma$ is the dissipation coefficient in the solvent and $D = k_BT/\gamma$ is the thermal diffusivity of the particle. Without active noise, $\xi_{act} = 0$, the particle is in thermal equilibrium, and the average barrier crossing time is the Kramers time, $\tau_K = \tau_r \exp(E_b/k_BT)$, where $\tau_r$ is the relaxation time within a single potential well [1, 37]. The various time scales used in this study are defined in Table 1. The active noise $\xi_{act}$ is generated from the discrete white noise $\xi_{PN}$, where each burst of random strength arrives at a discrete interval following Poisson distribution, using the active Ornstein–Uhlenbeck process (see SI for noise generation process):

$$\tau_c d\xi_{act}(t)/dt = -\xi_{act}(t) + \sqrt{2}\xi_{PN}(t). \quad (2)$$

Equation (2) generates exponentially correlated nonGaussian active noise with a zero mean $\langle \xi_{act} \rangle = 0$, and autocorrelation

$$\langle \xi_{act}(t)\xi_{act}(t') \rangle = f_{act}^2 \exp(-|t-t'|/\tau_c). \quad (3)$$

Here, $f_{act} \equiv \sqrt{C/(1+\tau_P/\Delta t)}$ characterizes the strength of active noise (the standard deviation of the active noise distribution), where $C$ is the variance of the Gaussian white noise from which the discrete white noise $\xi_{PN}$ is generated (Eq. (S9) in SI), $\tau_P$ is the average noise arrival time, and $\Delta t$ is the noise input interval (the sampling time). Typical noise arrival time distributions are shown in Fig. S2 in SI. Note that although each kick arrives at an average interval $\tau_P$, we want to stress that the strength and direction of the kicks are purely random

**Table 1.** List of timescales used in the study.

| | |
|---|---|
| $\Delta t$ | Sampling time |
| $\tau_K$ | Kramers time in a thermal bath |
| $\tau_r$ | Thermal relaxation time |
| $\tau_c$ | Active noise correlation time |
| $\tau_P$ | Average active noise arrival time |
| $\tau_d$ | Residence time |
| $\tau_{mod}$ | Modulation time for periodic modulation |

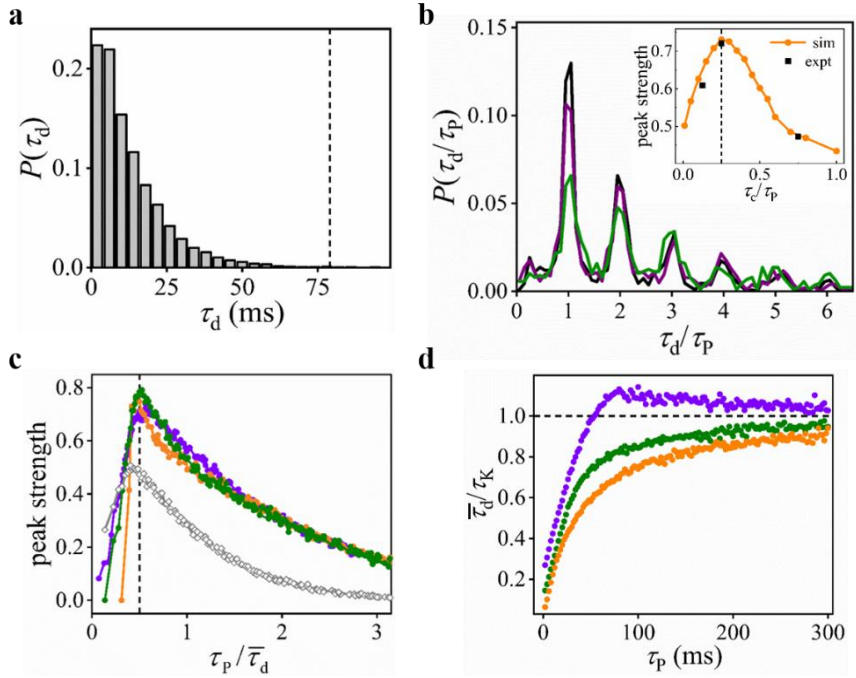

Fig. 2 (a) Experimentally measured residence time distribution (RTD) for $f_{act} \approx 0.5$ pN, $\tau_c \approx 0.7$ ms, and $\tau_P \approx 0$. Here, the average residence time $\bar{\tau}_d \approx 12.6 \pm 0.1$ ms is much less than the Kramers time, $\tau_K \approx 79$ ms (dashed vertical line). (b) Experimentally measured particle's RTD, where the residence time is normalized by noise arrival interval $\tau_d/\tau_P$, for the same $f_{act}$ and $\tau_P$ as in Fig. 1b with $\tau_c \approx 21$ ms (olive), 7 ms (black), and 3.5 ms (purple). Inset: Plot of the strength of the first peak as a function of $\tau_c/\tau_P$ for experimental data in the main panel (black squares) and obtained from the data obtained by solving Eq. (1)) numerically. The dashed vertical line denotes $\tau_c/\tau_P = 0.25$. (c) (Numerical result obtained by solving Eq. (1)) The strength of the first peak in the RTD for the particle as a function of $\tau_P/\bar{\tau}_d$ for fixed $\sqrt{C} \approx 20$ pN and $\tau_c \approx 1.25$ ms (orange), 7.5 ms (olive), and 20 ms (violet). The gray open circle is the plot of the second peak strength for $\tau_c \approx 7.5$ ms. The vertical dashed line corresponds to $\tau_P/\bar{\tau}_d = 0.5$. (d) Normalized average residence time as a function of $\tau_P$ for the like-coloured data in (c).

and follow white Gaussian noise (Fig. S1). Thus, the active noise acts on the system randomly.

The significance of our noise generation approach is that all three parameters, $C$, $\tau_c$, and $\tau_P$, can be independently varied. In particular, $\xi_{act}$ becomes AOU noise in the limit $\tau_P = 0$, and white Gaussian noise when both $\tau_c$ and $\tau_P$ vanish (Fig. S1). The power spectral density (PSD) of the nonGaussian active noise matches with the AOU noise of the same strength and correlation time (Fig. S1(f)). Thus, the nonGaussian active noise is similar to the AOU noise with reduced strength $\sqrt{C/(1+\tau_P/\Delta t)}$. A tracer particle in the presence of the nonGaussian active noise shows enhanced Gaussian or nonGaussian diffusion depending on the active noise parameters (see Eq. S12 and Fig. S5 in SI). The active noise thus mimics the real active baths such as the bath of swimming cells[17, 35] and cellular environments[32, 33, 38].

**Generating a double-well potential.** The double-well potential and the active noise in the experiment were generated simultaneously using the active optical feedback trap (AOFT) technique, an upgraded version of the optical feedback trap technique [39, 40]. The experimental setup is similar to the one used in [41-45] and also expounded in Fig. S3 in SI. To this end, a 2.0 $\mu$m diameter polystyrene particle suspended in deionized water at room temperature $T = 296 \pm 1$ K was trapped in a harmonic potential, $V_{op}(x,t) = (k/2)(x - x_c(t))^2$, generated by the trapping laser, where $x_c$ is the trap's center and $k$ is its stiffness.

The quadrant photodiode (QPD) measures the particle position ($x$ with respect to the trap center $x_c$) with high precision (~1 nm). The signal from the QPD is acquired by a field-programmable gate array (FPGA) card using a custom-written LabVIEW FPGA program at $\Delta t \approx 70$ $\mu$s sampling time. The FPGA computes the feedback force, $f_{DW} = -\partial_x V_{DW}(x)$, required for generating the double-well potential. The active noise is imposed by adding numerically-generated nonGaussian active noise, $\xi_{act}(t) \equiv -ky(t)$, to the feedback force $f_{DW}$. Equations (S1)-(S3) and Fig. S1 in the SI explain how $y(t)$ is generated. Each value of $y(t)$ – with Gaussian-distributed random amplitude $y$ of zero mean zero, $\langle y \rangle = 0$, and variance $\sigma^2$ – is randomly drawn from a Poisson distribution with an average interval $\tau_P$, and decays exponentially with correlation time $\tau_c$ (see Fig. 1a and Fig. S1b in the SI). The variance of the active burst in Eq. (2) is then $C/(1+\tau_P/\Delta t) = k^2\sigma^2/(1+\tau_P/\Delta t)$. The resultant force $f_{DW} + \xi_{act}$ is applied to the particle in the form of an optical feedback force by shifting the trap center instantaneously, using an acoustic optical deflector (AOD), to $x_c(t) = (1 + 4E_b/x_m^2 k)x(t) - (4E_b/x_m^4 k)x^3(t) + y(t)$. On repeating this protocol many times, the particle feels effective double-well potential and active noise. The AOFT technique thus generates the virtual double-well potential and active noise simultaneously, consequently the particle moves in the double-well potential mimicking real active baths. The maximum displacement of the particle that the QPD can measure such that the stiffness of the optical trap remains constant is ~ 0.6 $\mu$m. This limits the maximum strength of the active noise to $f_{act} \sim 0.5$ pN. In this study, we set $E_b = 3k_BT$ and $x_m = 50$ nm. In addition, the trap stiffness of the optical trap $k \approx 10$ pN/$\mu$m was obtained experimentally from the equipartition theorem [45]. The relaxation time of the particle in the optical trap is then given by $\tau_{op} = \gamma/k \approx 1.73$ ms.

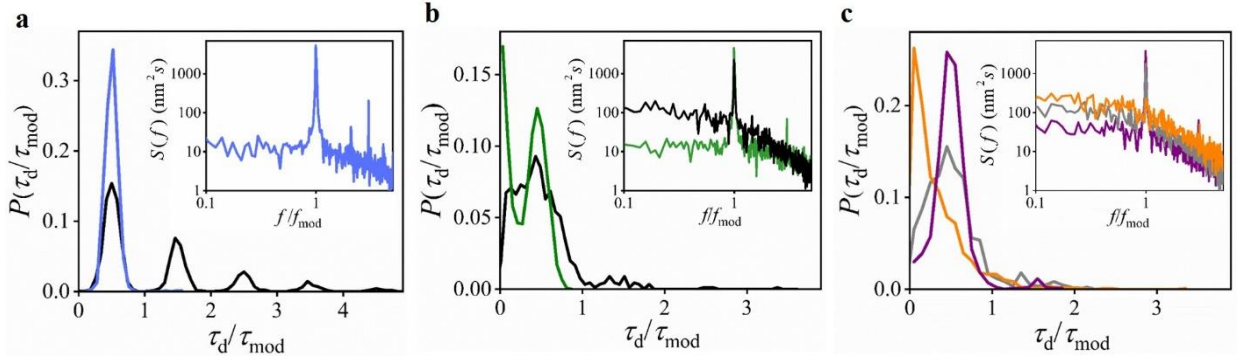

Fig. 3. Experimentally measured residence time distributions (RTDs) of the particle under the periodic modulation of the double-well potential. (a) In the absence of active noise with the modulation amplitude $A \approx 0.3$ pN and period $\tau_{\mathrm{mod}} \approx 0.38\tau_K$ (black) and $2\tau_K$ (blue). Inset: Power spectral density of the particle under the resonant condition $\tau_{\mathrm{mod}} \approx 2\tau_K$. (b) RTD under the resonant condition, $\tau_{\mathrm{mod}} = 2\tau_K$ and $A \approx 0.3$ pN, in the presence of Gaussian noise of strength $\sqrt{C} \approx 0.5$ pN and correlation time $\tau_c \approx 0.35$ ms (olive) and 35 ms (black). Inset: Corresponding power spectral densities. (c) RTD under the resonant condition, $\tau_{\mathrm{mod}} = 2\tau_K$ and $A \approx 0.3$ pN, in the presence of nonGaussian noise with fixed $\sqrt{C} \approx 1$ pN and $\tau_c \approx 35$ ms and $\tau_P \approx 3.5$ ms (orange), 35 ms (gray), and 350 ms (purple). Inset: Corresponding power spectral densities.

**NonGaussian distribution of particle position.** Figure 1b shows the PDFs of the particle position in the symmetric double-well potential, in the presence of nonGaussian active noise of fixed strength $f_{act} \approx 0.5$ pN (comparable to the typical strength of the active fluctuations in living cells [26, 32, 46]), and noise arrival time $\tau_P \approx 28$ ms ($> \tau_r \approx 4$ ms) obtained by analysis of the particle trajectories (see Fig. S4 a to c in SI). For $\tau_c < \tau_r$, the PDF exhibits two symmetric peaks centered around $\pm x_m$. The central region of the PDF is described by a Boltzmann distribution $P(x) \sim \exp(-U(x)/k_BT)$, where $U(x)$ is the effective double-well potential (see Fig. S4d in the SI), albeit with a reduced effective barrier height. However, when $\tau_c > \tau_r$ both the effective barrier height and well separation are larger than their values in a thermalized system (Fig S4e-f). Remarkably, each peak in Fig. 1b is always Gaussian near its center, but the outer tails are often nonGaussian. We found that the PDFs become nonGaussian only when $\tau_c \lesssim \tau_P$ and $f_{act} \gtrsim f_{th}$ where $f_{th} = (8k_BTE_b/x_m^2)^{1/2} \approx 0.4$ pN is the thermal force strength at the potential wells. This condition for obtaining a nonGaussian PDF is similar for diffusion in a simple harmonic potential (Fig. S5 in the SI).

**SR without periodic modulation.** To analyze the hopping dynamics of the particle in the symmetric double-well potential, we measured the probability distribution $P(\tau_d)$ of the residence time $\tau_d$, i.e., the time the particle remains within the potential well before hopping into the other potential well. Note that, for the particle hopping in the symmetric double-well potential in the thermal bath alone, i.e., in the absence of external periodic force or active noise, the average residence time $\bar{\tau}_d$ is equal to the Kramers time $\tau_K$. The residence time $\tau_d$ is determined based on the particle trajectories (as in [1, 9, 14]; see Fig. S7a-c in the SI for typical particle trajectories and active noise trajectories). In the absence of active noise, the RTD decays exponentially, with a mean residence time $\bar{\tau}_d$ equal to $\tau_K$ [1, 37, 47]. In the presence of active noise, in the regime $\tau_P < \tau_r$, the RTD remains exponential (Fig 2a), with $\bar{\tau}_d < \tau_K$.

However, for $\tau_P \gtrsim \tau_r$ and $f_{act} \gtrsim f_{th}$, the RTD displays a series of peaks, each centered at the integral multiple of $\tau_P$, i.e., $(\tau_d)_n = n\tau_P$ (Fig. 2b). The height of each peak decreases exponentially with its order $n$ (Fig. S6). In addition, the height of the first peak increases with the correlation time $\tau_c$ and assumes a maximum at a finite value of $\tau_c$. For further quantification, we measured the strength of the first peak (area under the first peak) $P_1 = \int_{(\tau_d)_1 - \tau_P/4}^{(\tau_d)_1 + \tau_P/4} P(\tau_d)d\tau_d$ [9, 48] (Fig. 2b inset and Fig. S7d) and found that it attains a maximum at $\tau_c \approx \tau_P/4$. The global maximum of $P_1$ is found when $\tau_r < \tau_c \lesssim 5\tau_r$ (Fig. S7d inset).

Furthermore, we measured the RTD of the particle as a function of $\tau_P$ while maintaining the active burst strength $\sqrt{C}$ and the correlation time $\tau_c$ constant. Similar to the above, a series of exponentially-decaying peaks centered at $n\tau_P$ are evident for $\tau_P \gtrsim \tau_r$. For a given $\tau_c$, the strength of the first peak is maximal when $\tau_P \approx \bar{\tau}_d/2$ (Fig. 2c). Note that for a fixed $\sqrt{C}$, a change in the noise arrival interval changes the strength of active noise $f_{act} \equiv \sqrt{C/(1+\tau_P/\Delta t)}$. Thus, we found that maximal synchronization can be achieved between the particle residence time and the active noise arrival interval by appropriately selecting the active noise parameters ($\tau_c$, $\tau_P$, or $f_{act}$). The pronounced maximum of the first peak strength demonstrates *bona fide* SR [1, 14, 49]. Importantly, the SR observed here is generated solely by the nonGaussian active noise, without any periodic modulation of the double-well potential. For $\tau_P \gg \tau_K$, the RTD again shows exponentially-decaying behavior because the kicking events are rare, and the mean residence time $\bar{\tau}_d$ saturates to the Kramers time $\tau_K$, as shown in Fig. 2d.

Figure 2d also shows the barrier crossing is enhanced or diminished depending on the active noise time scales ($\tau_c$ and $\tau_P$).

We found that in presence of active noise $\bar{\tau}_d$ is generally less than $\tau_K$. However, for strongly correlated active bath with $\tau_P \gtrsim \tau_c \gg \tau_r$ (see violet curve in Fig. 2d), $\bar{\tau}_d > \tau_K$, leading to the slowdown of the barrier crossing. Also, a recent study on the particle motion in the double-well potential in the presence of viscoelastic bath demonstrated the barrier-crossing enhancement is controlled by the time scales associated with the viscoelastic bath[50].

The observed barrier crossing enhancement and particle synchronization in the presence of nonGaussian active noise can be intuitively explained: The active noise randomly injects energy into the system, with a mean interval $\tau_P$, and each pulse decays with a correlation time $\tau_c$. For $\tau_c$ and $\tau_P < \tau_r$, several bursts kick the particle during its thermal relaxation time, thus increasing its effective temperature and enhancing the barrier crossing rate. In Fig. S7 we analyze the particle trajectory under the influence of the active noise of fixed $f_{act} \approx 0.5$ pN and $\tau_P/\tau_r \gg 1$ with different $\tau_c$. For $\tau_P/\tau_r \gg 1$ and $\tau_c/\tau_r \ll 1$ (Fig. S7a), each active burst decays faster than $\tau_r$. The active bursts drive the particle through its short correlation time $\tau_c \ll \tau_r$ and cease to act on the particle during the time interval $\tau_P - \tau_c$, allowing to recover the thermal condition (Fig. 1a). Consequently, the barrier crossing is less synchronized with the active noise. On the other hand, for $\tau_P/\tau_r \gg 1$ and $\tau_P/\tau_c < 1$ (Fig. S7c), many correlated active bursts arrive during the correlation time $\tau_c \gg \tau_r$ and the active noise becomes AOU noise. In the presence of such AOU noise with long correlation time, the particle is driven in the same direction for $\tau_c \gg \tau_r$, thereby slowing down the barrier crossing.

Maximum synchronization between the barrier crossing and the active noise is observed when $\tau_P/\tau_r \gg 1$ and $\tau_r < \tau_c < \tau_P$, (Fig S7b). Here, the strength of individual kick is large enough for driving the particle to cross the barrier. Moreover, each active burst decays fully before another burst arrives (the active bursts are not correlated with each other). Thus, a particle, for example, in the left well (see Fig. 1a) may diffuse to the right well when acted by the positive active burst with correlation duration $\tau_c > \tau_r$ and relaxes in thermal equilibrium during the time $\tau_P - \tau_c$ inside the right well. If the particle does not cross the barrier, it waits for the duration $\tau_P - \tau_c$ during which the particle relaxes in thermal equilibrium inside the left well. Thus, nonGaussian noise randomly modulates the barrier height and potential-well separation, with an average modulation period $\tau_P$ (Fig. 1a). The ideal timing for the particle to cross the barrier is when its height is the lowest. This optimality condition can be achieved by tuning either $\tau_P$, which controls the noise arrival interval and the noise strength $f_{act} \equiv \sqrt{C/(1+\tau_P/\Delta t)}$, or $\tau_c$, which controls the decay time of the active bursts and the thermal relaxation duration $\tau_P - \tau_c$ of the particle. Thus, we see that the resonance condition, $\tau_r < \tau_c \lesssim 5\tau_r$ and $4\tau_c \approx \tau_P \approx \bar{\tau}_d/2$, can be achieved by varying these noise parameters.

**Recovery of generic SR by nonGaussian active noise.** To gain further insight, we also studied the generic form of SR experimentally with periodic modulation of the double-well potential in the presence of nonGaussian active noise by using the AOFT technique. To this end, the double-well potential is periodically tilted with an amplitude $A$ and a period $\tau_{mod}$, $V(x,t) = V_{DW}(x) - Ax\sin(2\pi t/\tau_{mod})$. In the absence of active noise, and for modulation times shorter than the Kramers time, $\tau_{mod} < \tau_K$, the RTD shows a series of peaks centered at odd multiples of $\tau_{mod}/2$ (black curve in Fig. 3a). On increasing the modulation time toward the resonant condition $\tau_{mod} \approx 2\tau_K$, the barrier crossing rate of the particle becomes synchronized with the modulation period, and a single peak centered at $\tau_{mod}/2$ is observed (blue curve in Fig. 3a). The PDF of the particle under the periodic modulation with $\tau_{mod} \approx 2\tau_K$ coincides with the thermal PDF (gray curve in Fig. 1b). The SR phenomenon under periodic forcing can also be identified through the power spectrum density (PSD) of the particle fluctuations [1]. A sharp peak is observed at the modulation frequency and a weak peak at the third harmonic (Fig. 3a inset). Thus, our optical feedback trap method can precisely measure SR in a thermal bath under periodic forcing. Compared to the previous experimental works [47, 49], which studied the SR of Brownian particles in double-well potentials with inter-well separation greater than $1\,\mu m$, we demonstrated here SR in a nanoscale double-well potential well separated by $2x_m = 100$ nm.

In the presence of active noise at $\tau_P \approx 0$, corresponding to the Gaussian regime of the active bath, we observed suppression of the SR: If the double-well potential is modulated sinusoidally with a period $\tau_{mod} \approx 2\tau_K$, the intensities of the resonant peaks in the RTD as well as PSD decrease as the correlation time $\tau_c$ increases (Fig. 3b and inset). The peaks disappear completely when $\tau_c \gg \tau_r$, and the active noise is stronger than the thermal noise. Our experimental observation agrees with the theoretical prediction in [28]. However, in the nonGaussian regime of the active bath, the resonant peak reappears at finite $\tau_P$, as shown in Fig. 3c. Likewise, a sharp peak at the modulation frequency is observed in the PSD (Fig. 3c inset). The resonant peak height increases with $\tau_P$ and recovers back to the purely thermal level when $\tau_P \gg \tau_K$, even when the active noise is stronger than the thermal noise, $f_{act} \gtrsim f_{th}$ (Fig. S8 in the SI). Thus, we recovered SR in the nonGaussian active bath under periodic forcing.

SR recovery in the presence of nonGaussian active noise at finite non-zero intervals, $\tau_P > 0$, can be explained as follows: For $\tau_P \approx 0$ (the Gaussian regime), the active noise supplies energy to the particle continuously and in a random direction for an average time $\tau_c$. Therefore, this noise might counteract the barrier crossing process, with its typical timescale $\tau_K$ and the tilting with its period $\tau_{mod}$, thereby suppressing the resonant behavior for long correlations $\tau_c$ ($\tau_c \gg \tau_r$). However, when $\tau_P > \tau_c \gg \tau_r$ (the nonGaussian regime), the noise strength decreases as $f_{act} \sim (1+\tau_P/\Delta t)^{-1/2}$. Furthermore, each active pulse completely decays before the arrival of another one; consequently, the particle is free of the active noise during the time interval $\tau_P - \tau_c$ (as in Fig. 1a) and becomes equilibrated by dissipating energy into the thermal bath. Thus, the thermal SR condition is recovered. This is evident from our observation that the central region of the effective potential—for which the PSD shows a sharp resonant peak (Fig. S8)—fits well to a thermally-activated potential (Fig. S8 inset).

## Conclusions

To sum, we studied the dynamics of a colloidal particle in a symmetric double-well potential in the presence of exponentially-correlated nonGaussian active noise. The RTD exhibited a series of peaks at integral multiples of the noise arrival interval. The strength of the first peak was maximized by either changing the correlation time or the noise arrival interval, demonstrating SR in the Brownian system without symmetry breaking. The generic form of SR with periodic modulation of the double-well potential, which deteriorates with Gaussian correlated noise, was recovered with nonGaussian correlated noise having a large noise arrival interval. The findings of this study indicate that nonGaussian active fluctuations may lead to the synchronization of various biomolecular processes, such as protein folding, enzymatic reactions, and signal transduction inside living cells.

## Author Contributions

## Conflicts of interest

## Acknowledgements

This work was supported by the Institute for Basic Science (Grant no. IBS-R020-D1). We thank Won Kyu Kim for helpful discussions.

# Supplementary Information

## *Bona fide* stochastic resonance under nonGaussian active fluctuations

Govind Paneru[ab], Tsvi Tlusty*[ab], and Hyuk Kyu Pak*[ab]

[1]Center for Soft and Living Matter, Institute for Basic Science (IBS), Ulsan 44919, Republic of Korea

[2]Department of Physics, Ulsan National Institute of Science and Technology, Ulsan 44919, Republic of Korea

## Active noise generation procedure

The active noise in the main text is generated numerically as follows: Let $Q_n$ be the sequence of identically distributed random numbers that follow a Gaussian distribution with variance $\sigma^2$,

$$P(Q) = \frac{1}{\sqrt{2\pi\sigma^2}} \exp(-\frac{Q^2}{2\sigma^2}). \tag{S1}$$

From these random numbers, we can generate a sequence of random numbers that follow a nonGaussian distribution (Fig. S1 (a)):

$$q_P(t) = \sum_{i=1} Q_i \delta(t-t_i). \tag{S2}$$

Here, each kick of duration $\Delta t \to 0$ arrives at a random interval $t_i$ following a Poisson distribution. The kick arrival time (time interval between successive kicks) thus follows the Poisson distribution with mean arrival time $\tau_P = \overline{t_{i+1} - t_i}$. Eq. S2 shows all random numbers of the sequence $Q_n$ from white Gaussian noise becomes zero except one arriving at the *i*th interval $t_i$; hence, the variance of the nonGaussian white noise $q_P$ is given by $\sigma^2/(1+\tau_P/\Delta t)$. The exponentially correlated nonGaussian noise $y(t)$ of correlation time $\tau_c$ can be generated recursively using the following relation [51],

$$y(t+\Delta t) = \exp(-\Delta t/\tau_c)\ y(t) + \sqrt{1-\exp(-2\Delta t/\tau_c)}\ q_P(t+\Delta t). \tag{S3}$$

The time traces of the exponentially correlated nonGaussian active noise $y(t)$ are shown in Fig. S1(b). The active force $\xi_{act}(t)$ is obtained by multiplying $y(t)$ with the trap stiffness $k$ as

$$\xi_{act}(t) = ky(t). \tag{S4}$$

The above noise generation approach is equivalent to the active Ornstein–Uhlenbeck process[52]:

$$\tau_c d\xi_{act}(t)/dt = -\xi_{act}(t) + \sqrt{2}\xi_{PN}(t), \tag{S5}$$

where $\xi_{PN}(t)$ is the nonGaussian white noise with a zero mean $\langle \xi_{PN}(t) \rangle = 0$ and correlation $\langle \xi_{PN}(t)\xi_{PN}(t') \rangle = [A/(1+\tau_P/\Delta t)]\delta(t-t')$; and A is the energy scale of the active force. The Kurtosis $K(t) = <[\xi(t)-<\xi(t)>]^4>/<[\xi(t)-<\xi(t)>]^2>^2$, which measures the degree of nonGaussianity of the probability distributions, of nonGaussian white noise $\xi_{PN}(t)$ and the exponentially correlated active noise $\xi_{act}(t)$, are shown in

Fig. S1(d) and Fig. S1(e), respectively. The active noise $\xi_{act}(t)$ becomes nonGaussian ($K>3$) for $\tau_P/\tau_c \gtrsim 0.2$. Integrating Eq. (S5), we get the formal solution for the nonGaussian active noise:

$$\xi_{act}(t) = \xi_{act}(0)e^{-t/\tau_c} + \frac{\sqrt{2}}{\tau_c}\int_0^t e^{-(t-t')/\tau_c}\xi_{PN}(t')dt'. \tag{S6}$$

From Eq. (S6), we can derive the autocorrelation of the nonGaussian active noise as

$$\langle \xi_{act}(t)\xi_{act}(t')\rangle = \langle (\xi_{act}(0))^2\rangle e^{-(t+t')/\tau_c} + \frac{2}{\tau_c^2}\int_0^t ds\, e^{-(t-s)/\tau_c}\int_0^{t'} ds'\, e^{-(t'-s')/\tau_c}\langle \xi_{PN}(s)\xi_{PN}(s')\rangle. \tag{S7}$$

By using $\langle \xi_{PN}(s)\xi_{PN}(s')\rangle = [A/(1+\tau_P/\Delta t)]\delta(s-s')$, Eq. S7 becomes

$$\langle \xi_{act}(t)\xi_{act}(t')\rangle = \langle (\xi_{act}(0))^2\rangle e^{-(t+t')/\tau_c} + \frac{A/(1+\tau_P/\Delta t)}{\tau_c}[e^{(t-t')/\tau_c} - e^{-(t+t')/\tau_c}]. \tag{S8}$$

In steady state, $\langle (\xi_{act}(0))^2\rangle$ is equal to the variance of the active noise, $\langle (\xi_{act}(t))^2\rangle = A/[\tau_c(1+\tau_P/\Delta t)]$. Hence, the autocorrelation of the nonGaussian active noise in steady state is given by

$$\langle \xi_{act}(t)\xi_{act}(t')\rangle = \frac{A/(1+\tau_P/\Delta t)}{\tau_c}e^{-|t-t'|/\tau_c}. \tag{S9}$$

The active force strength $f_{act} \equiv \sqrt{C/(1+\tau_P/\Delta t)}$ of the nonGaussian active noise is related to A and $\tau_c$ as $f_{act} = \sqrt{A/[\tau_c(1+\tau_P/\Delta t)]}$. For $\tau_P \to 0$ and $\Delta t \to 0$, $A/(1+\tau_P/\Delta t) \to A$, hence $\xi_{PN}(t)$ reduces to the Gaussian white noise $\xi_{GN}$ of zero mean and correlation $\langle \xi_{GN}(t)\xi_{GN}(t')\rangle = A\delta(t-t')$ and the active noise $\xi_{act}(t)$ in Eq. (S5) reduces to the active Ornstein–Uhlenbeck noise with correlation $\langle \xi(t)\xi(t')\rangle_{OU} = (A/\tau_c)e^{-|t-t'|/\tau_c}$.

Solving Eq. (S5) by the Fourier transform method, we can derive an expression for the power spectral density (PSD) of the nonGaussian active noise:

$$S(\omega) = \frac{2f_{act}^2 \tau_c}{1+\tau_c^2\omega^2}. \tag{S10}$$

Figure S1(f) shows the PSD of the nonGaussian active noise agrees well with Eq. (S10).

**Particle dynamics in a harmonic potential**

We consider the one-dimensional motion of the particle in a harmonic potential $V(x) = (1/2)kx^2$ in an active bath. The motion of the particle is given by the overdamped Langevin equation:

$$\gamma\frac{dx}{dt} = -kx + \xi_{th} + \xi_{act}. \tag{S11}$$

The mean squared displacement (MSD) of the particle in the harmonic potential during the time interval $t$ in steady-state can be calculated by solving Eq. (S11) by Laplace transform method [22]:

$$\langle \Delta x^2(t) \rangle = 2D\tau_r \left(1-e^{-t/\tau_r}\right) + \frac{2A\tau_r}{\gamma^2(1+\tau_P/\Delta t)} \frac{\left[1-e^{-t/\tau_r} - (\tau_c/\tau_r)(1-e^{-t/\tau_c})\right]}{[1-(\tau_c/\tau_r)^2]}. \tag{S12}$$

Figure S5(c) shows that the MSD of the particle in the presence of nonGaussian active noise fits well with Eq. (S12). Also, we show that the MSD with nonGaussian active noise coincides with the MSD of the AOU noise of the same $f_{act}$ and $\tau_c$. Thus, the steady state dynamics in the presence of nonGaussian active noise is similar to that of the AOU noise.

## Figures

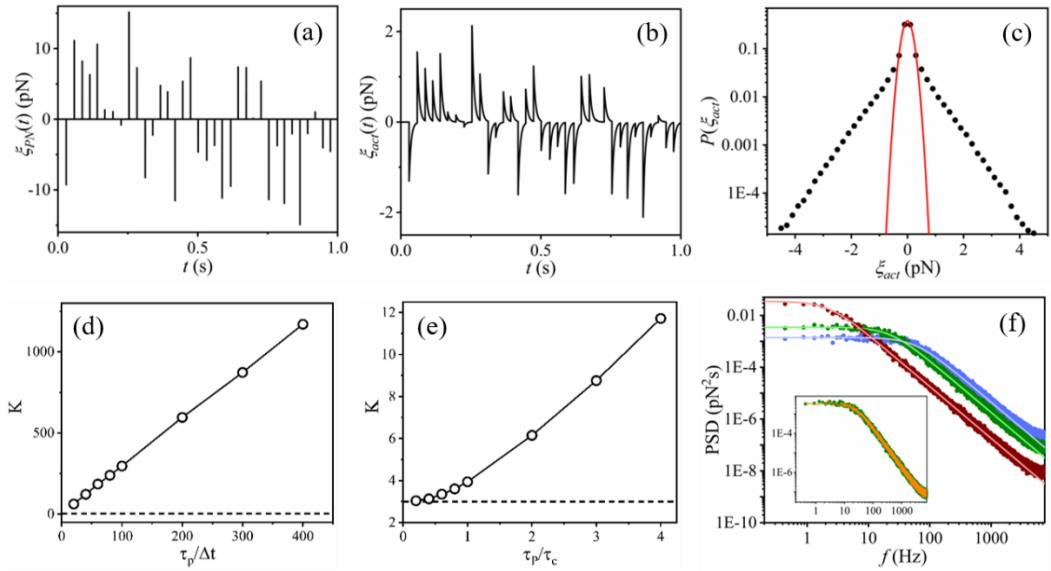

**Fig. S4**. (a) Trace of the nonGaussian white noise $\xi_{PN}(t)$ with $\tau_P = 28$ ms generated from Gaussian white noise of variance $C = k^2\sigma^2 = (10 \text{ pN})^2$, following Eq. (S2). The variance of the nonGaussian white noise is given by $C/(1+\tau_P/\Delta t) \approx (0.5 \text{ pN})^2$. The average waiting time between the kicks is $\tau_P$. Note that although each kick arrives at an average interval $\tau_P$, the strength and direction of the kicks are purely random, which follow white Gaussian noise. (b) The exponentially correlated nonGaussian active noise $\xi_{act}(t)$ of strength $f_{act} \equiv \sqrt{C/(1+\tau_P/\Delta t)} \approx 0.5$ pN with $\tau_P = 28$ ms and $\tau_c = 7$ ms generated from the nonGaussian white noise in panel (a) using Eq. (S4). (c) Probability distribution function (PDF) of the nonGaussian active noise in panel (b). The solid red curve is the Gaussian fitting. (d) Kurtosis of the nonGaussian white noise as a function of $\tau_P/\Delta t$. (e) Kurtosis of nonGaussian active noise as a function of $\tau_P/\tau_c$. The horizontal dashed line in (d) and (e) correspond to Gaussian distribution $K = 3$. (f) Power spectral density (PSD) of the nonGaussian active noise of fixed $f_{act} = 0.5$ pN and $\tau_P = 28$ ms with $\tau_c = 2.8$ ms (blue), 7 ms (olive), and 70 ms (wine). The solid curves are the theoretical plots using Eq. (S10). Inset: PSD of the nonGaussian active noise of $f_{act} = 0.5$ pN, $\tau_P = 28$ ms, and $\tau_c = 7$ ms (olive) coincides with the active Ornstein–Uhlenbeck noise of same strength $f_{act} = 0.5$ pN and correlation time $\tau_c = 7$ ms (orange).

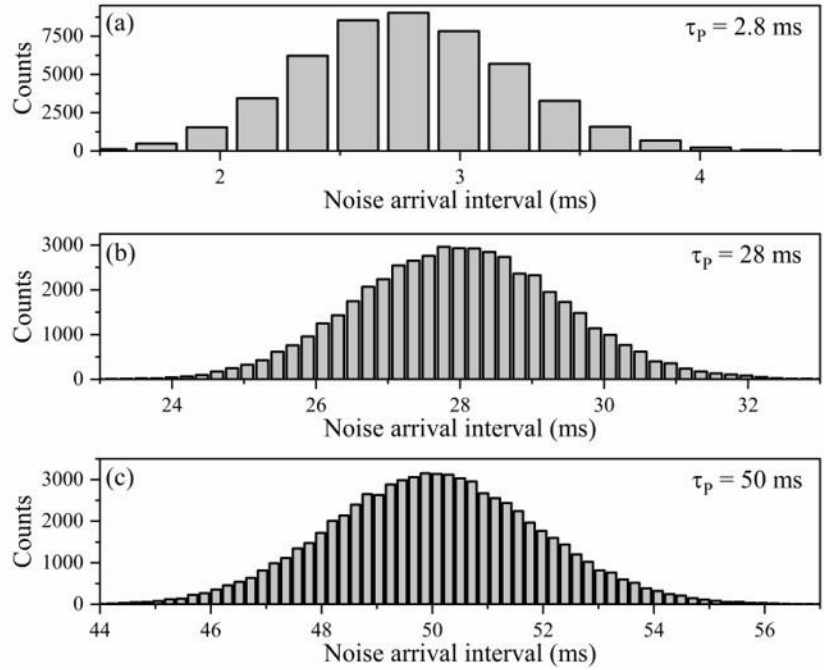

**Fig. S5**. Noise arrival time distribution for nonGaussian active noise follows a Poisson distribution with average noise arrival time (a) $\tau_P = (2.8 \pm 0.44)$ ms, (b) $\tau_P = (28 \pm 1.41)$ ms, and (c) $\tau_P = (50 \pm 1.88)$ ms.

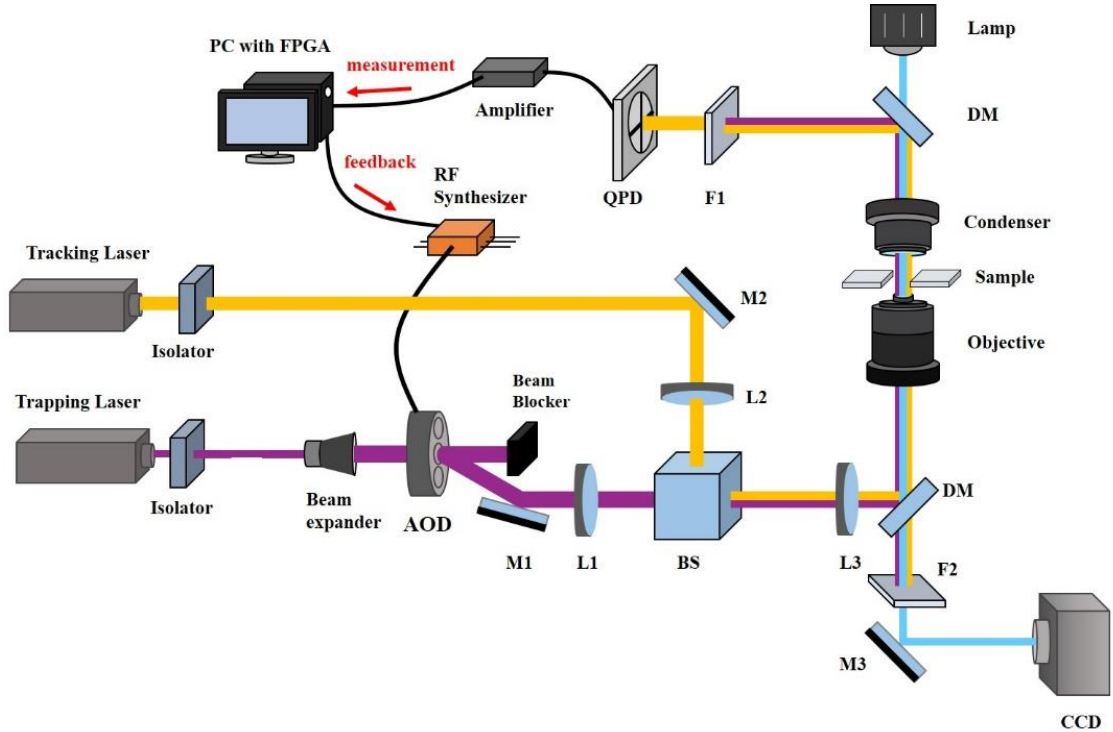

**Fig. S6**. Schematics of the active optical feedback trap (AOFT) setup. M1, M2, M3: mirror. L1, L2, L3: Lens. BS: beam splitter, DM: dichroic mirror, F1, F2: filter. CCD: camera.

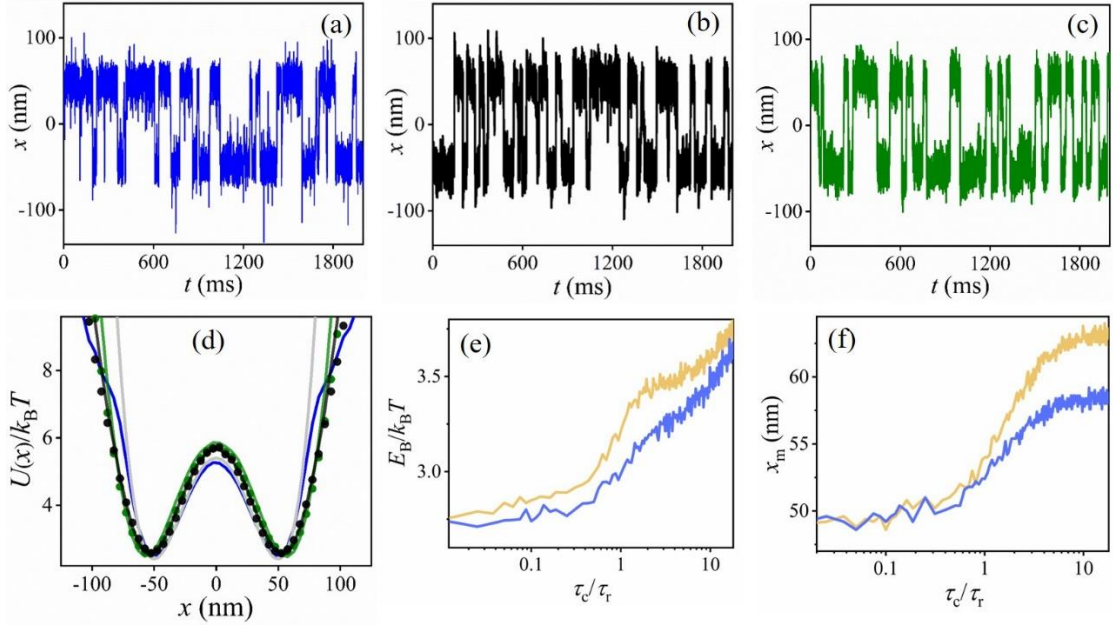

**Fig. S7**. Trajectories of the particle in a double-well potential, $E_b/k_BT = 3$ and $x_m = 50$ nm (top panels), in the presence of nonGaussian active noise of fixed strength $f_{act} \approx 0.5$ pN and average noise arrival time $\tau_P \approx 28$ ms with correlation time (bottom panels). (a) $\tau_c \approx 0.28$ ms (blue, numerical result), (b) 7 ms (black), and (c) 21 ms (green). The PDFs of the particle position in Fig. 1b of the main text are from these trajectories. (d) Effective double-well potentials $U(x)/k_BT \sim -\ln P(x)$ for the like-colored data in panels (a) to (c) (also PDFs in Fig. 1b in the main text). The dotted curves represent the experimental results. The solid curves are obtained from the numerical simulation of Eq. (1) in the main text. The gray curve is the theoretical plot of the symmetric double-well potential with $E_b/k_BT = 3$ and $x_m = 50$ nm. (Numerical result) (e)-(f) Plot showing the variation of the barrier height $E_b/k_BT$ and well separation $x_m$ as a function of $\tau_c/\tau_r$ for the particle in symmetric double-well with $E_b/k_BT = 3$ and $x_m = 50$ nm in the presence of the nonGaussian active noise with $\tau_P = 20$ ms and $f_{act} = 0.5$ pN (blue) and $f_{act} = 1$ pN (orange).

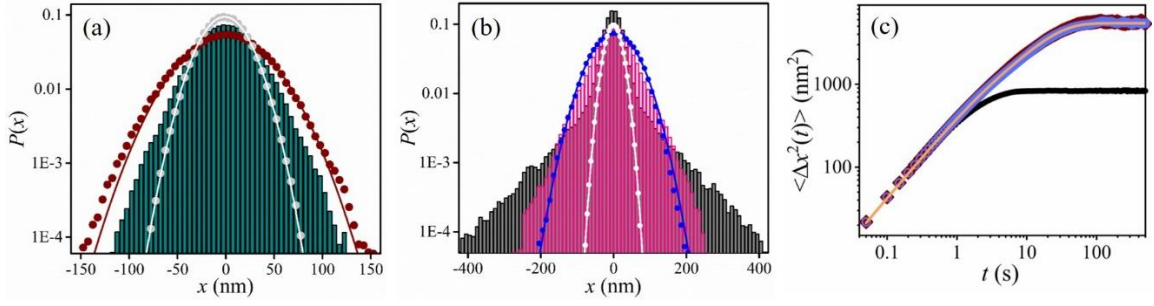

**Fig. S8**. PDF of the particle position in harmonic potential $V_{op}(x) = (k/2)(x-x_c)^2$ in the presence of nonGaussian active noise. Here, the active noise is injected into the particle in the form of feedback force $f_{act}(t) = -ky(t)$ that corresponds to the shift of the potential center by $x_c(t) = x(t) - y(t)$. (a) (experimental result) PDF of the particle position in the harmonic potential of stiffness $k \approx 9.1 \text{ pN}\mu\text{m}^{-1}$ in the thermal bath (gray circles), in the presence of non-Gaussian noise of strength $\sqrt{C} = 4.6$ pN and correlation time $\tau_c = 17.5$ ms with noise arrival interval $\tau_P = 14$ ms (wine circles), and 35 ms (dark cyan bars). The solid curves are the Gaussian fittings. (b) (numerical result obtained by solving Eq. S11) PDF of the particle position in the harmonic potential in the thermal bath (white circles), in the presence of non-Gaussian noise of fixed strength $f_{act} = 0.5$ pN and correlation time $\tau_c = 25$ ms with $\tau_P = 5$ ms (blue circles), 50 ms (pink bars), and 250 ms (black bars). The solid curves are the Gaussian fittings. We found that the PDFs are non-Gaussian only when $\tau_c \lesssim \tau_P$ and $f_{act} \gtrsim f_{th}$, where $f_{th} = \sqrt{k_BTk} \approx 0.2$ pN is the thermal strength. (c) Mean squared displacement of a particle in the harmonic potential in steady-state in the presence of the nonGaussian active noise of $f_{act} = 0.5$ pN, $\tau_c = 20$ ms, and $\tau_P = 40$ ms (light blue) coincides with the MSD in the presence of AOU noise of same $f_{act} = 0.5$ pN and $\tau_c = 20$ ms (wine). The orange solid curve is the plot of Eq. (S12). The black data is the MSD in the thermal bath.

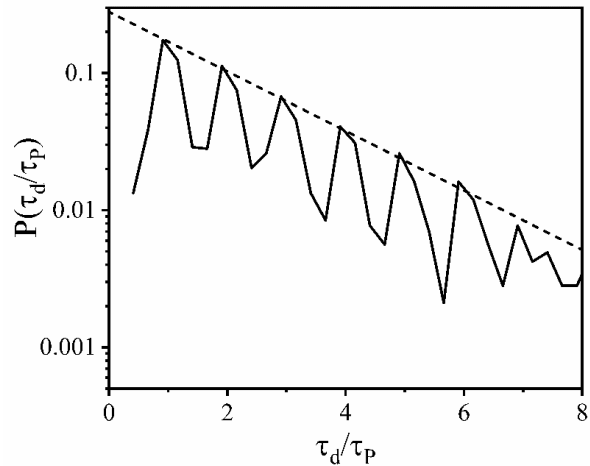

**Fig. S9.** Semi-log plot of the residence time distribution of the particle in the symmetric double-well potential in the presence of nonGaussian active noise of $f_{act} \approx 1$ pN and $\tau_c/\tau_P \approx 0.25$. The dotted line corresponds to $y = 0.28\exp(-0.5x)$. The fitting of the dotted line with the peak heights demonstrates that the height of each peak decreases exponentially with their order n.

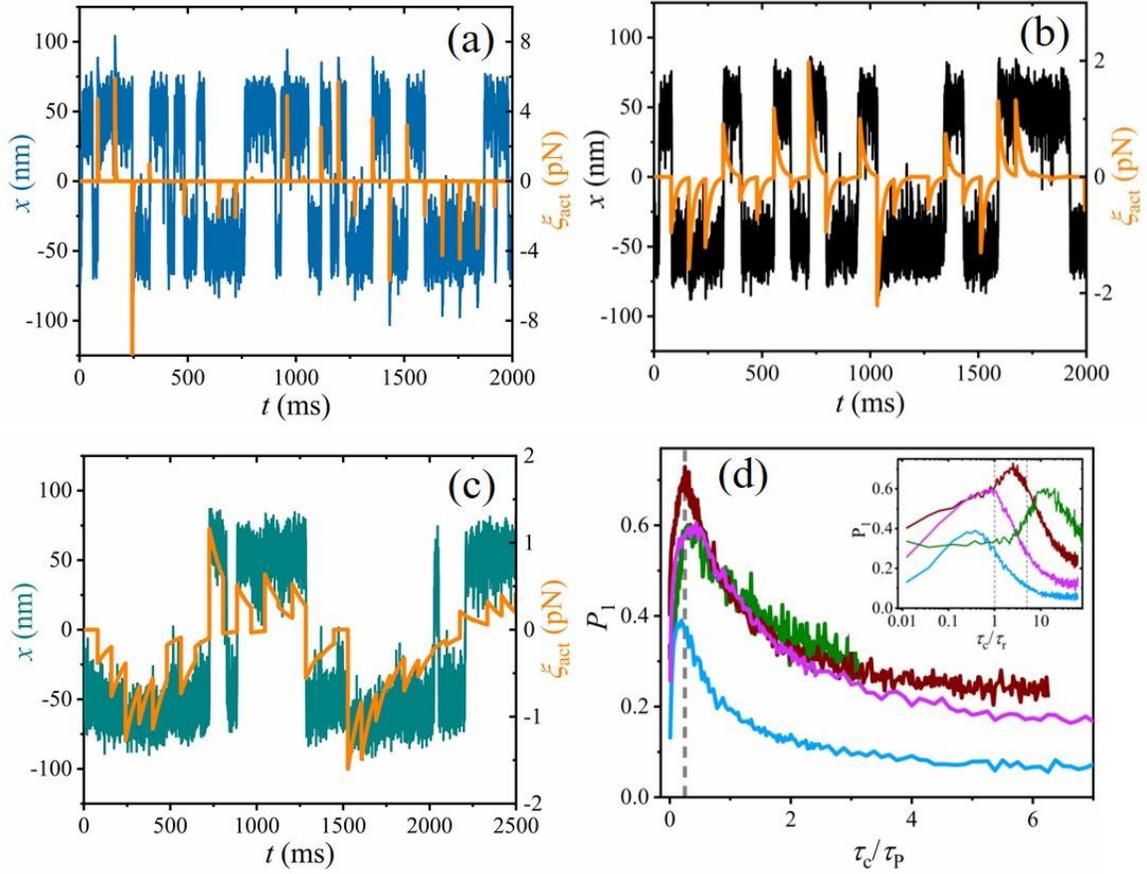

**Fig. S10**. (Numerical result obtained by solving Eq. (1) in the main text) (a) to (c) Particle trajectories and nonGaussian active noise trajectories (orange curves) for the motion of the particle in the double well potential in the presence of nonGaussian active noise of fixed $f_{act} \approx 0.5$ pN and $\tau_P/\tau_r \approx 20$ with (a) $\tau_c/\tau_r \approx 0.5$, (b) $\tau_c/\tau_r \approx 5$, and (c) $\tau_c/\tau_r \approx 30$. (d) Plot of the first peak strength $P_1$ as a function of $\tau_c/\tau_P$ for fixed $f_{act} \approx 0.5$ pN and $\tau_P \approx 8$ ms (light blue) $\tau_P \approx 16$ ms (light purple), $\tau_P \approx 40$ ms (wine), and $\tau_P \approx 160$ ms (olive). For all cases, $P_1$ is maximum near $\tau_c/\tau_P \approx 0.25$ (dashed vertical line). Inset: Same data in the main panel plotted as a function of $\tau_c/\tau_r$ shows the global maximum of $P_1$ is when $\tau_r < \tau_c \lesssim 5\tau_r$.

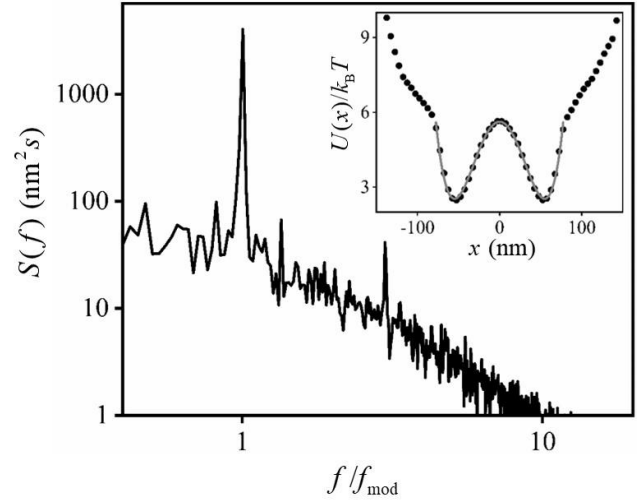

**Fig. S11**. (Numerical result obtained by solving Eq. (1) in the main text) Power spectral density in the presence of the nonGaussian active noise of strength $f_{act} \approx 1$ pN (greater than thermal strength $f_{th} \approx 0.4$ pN) with $\tau_c \approx 25$ ms and $\tau_p \approx 1000$ ms (non-Gaussian regime), under the same resonant condition of Fig. 3(c) in the main text. Inset: the effective double-well potential (black circles) for the same data in the main panel. The gray solid curve fits well with the symmetric double-well potential $V_{DW}(x)$ of $E_b/k_BT = 3$ and $x_m = 50$ nm with the fitting parameter $E_b/k_BT = 3.09 \pm 0.06$ and $x_m = 54.9 \pm 0.2$ nm, showing that the effective double-well potential follows Gaussian distribution near the center with non-Gaussian outer tails.